# Stable nonlinear amplification of solitons without gain saturation

Olga V. Borovkova,[1] Valery E. Lobanov,[1] and Boris A. Malomed[2,1]
[1]*ICFO-Institut de Ciencies Fotoniques, and Universitat Politecnica de Catalunya, Mediterranean Technology Park, 08860 Castelldefels (Barcelona), Spain*
[2]*Department of Physical Electronics, School of Electrical Engineering, Faculty of Engineering, Tel Aviv University, Tel Aviv 69978, Israel*

**Abstract**

We demonstrate that the cubic gain applied in a localized region, which is embedded into a bulk waveguide with the cubic-quintic nonlinearity and uniform linear losses, supports stable spatial solitons in the absence of the quintic dissipation. The system, featuring the bistability between the solitons and zero state (which are separated by a family of unstable solitons), may be used as a nonlinear amplifier for optical and plasmonic solitons, which, on the contrary to previously known settings, does not require gain saturation. The results are obtained in an analytical form and corroborated by the numerical analysis.

PACS 42.65.Tg; 05.45.Yv; 42.60.Da; 47.54.-r

*Introduction.* Self-trapping of spatial solitons is one of the fundamental topics of nonlinear optics [1-3]. In many realizations, waveguides carrying the solitons are used as laser cavities, that makes it necessary to include the loss and compensating gain into the analysis. This gives rise to models combining the transverse diffraction and Kerr nonlinearity with the gain and loss terms, in the form of the respective complex Ginzburg-Landau (CGL) equations. These models support spatial dissipative solitons through the concurrent self-focusing-diffraction and gain-loss balances [4].

The stability of the zero background is an obvious condition necessary for the creation of stable dissipative solitons, which rules out the simplest model based on the single-component CGL equation with the spatially uniform linear gain (dissipative solitons may be made stable in a system of linearly coupled equations modeling dual-core systems, with the linear gain and loss acting in different cores [5-7]). Instead, stable solitons are produced by the CGL equation of the cubic-quintic (CQ) type, which combines linear and quintic loss terms and the cubic gain [8-12]. The CQ-CGL equations of the same type serve as models of nonlinear optical amplifiers, which were designed with the objective to reshape solitons without amplifying the ambient noise [13-16].

Another possibility was recently elaborated within the framework of the single cubic CGL equation, where the linear gain is applied at a "hot spot" (HS), i.e., a locally pumped region created in a lossy waveguide [17,18], or at several HSs [19-22]. This setting can be built using an appropriate profile of the concentration of gain-inducing dopants, or illuminating a uniformly doped waveguide by a properly focused pump beam. In this way, stable dissipative solitons may be supported by the balance between the localized amplification and bulk dissipation. In particular, in the limit case of the $\delta$-functional HSs, solutions for the pinned dissipative solitons can be found in an exact analytical form [17,19,21]. In models with finite-width HSs, soliton solutions were found in a numerical form [17-22], including two-dimensional vortices supported by the gain applied along a ring [23,24]. Earlier, a similar model was introduced for gap solitons pinned to the HS inserted into a lossy Bragg grating [25]; more recently, a similar setting was developed for lasing media [26].

A natural issue is a possibility to design an amplifier with localized *cubic* gain and uniformly distributed linear loss, with the aim to provide clean amplification of the solitons, avoiding concomitant raise of the surrounding noise. A challenging aspect of such a setting is securing the self-trapping of stable solitons pinned to the cubic-gain HS, *without* postulating the presence of the overall-stabilizing quintic dissipation. Dissipative solitons in the CGL equation combining the uniformly distributed linear loss and cubic gain, without the quintic loss, are unstable against the blowup (some "exotic" quasi-periodic spatial modes, but not solitons, may be stabilized if the equation includes the linear diffusion term [27]). In this work, we demonstrate, by means of analytical and numerical methods, that *stable* spatial solitons can be supported by the localized cubic gain competing with the uniform linear loss, if quintic terms are represented solely by the uniform self-defocusing added to the cubic self-focusing, while the quintic loss is absent. The realization of the model may be provided by the same settings which were used in previously studied schemes of the nonlinear amplification in optics [13-16] and plasmonics [28]. In particular, a straightforward possibility is to use a second-harmonic-generating element, in which the amplification is provided in the second harmonic, while the input and output signals are carried by the fundamental frequency [16].

*Analytical results*. The propagation of the electromagnetic wave in a medium featuring, as said above, the focusing cubic and defocusing quintic nonlinearities, linear loss, and localized cubic gain, is described by the CGL equation for amplitude $q$ of the optical or plasmonic field:

$$i\frac{\partial q}{\partial \xi}=-\frac{1}{2}\frac{\partial^2 q}{\partial \eta^2}-|q|^2 q+\varepsilon_5|q|^4 q-i\gamma q+i\Gamma\exp\left(-\frac{\eta^2}{w_\Gamma^2}\right)|q|^2 q. \tag{1}$$

Here the transverse coordinate, $\eta$, and propagation distance, $\xi$, are normalized to the characteristic transverse scale and diffraction length, respectively, $\gamma$ is the linear-loss coefficient, $\Gamma$ and $w_\Gamma$ are the strength and width of the cubic-gain HS, and $\varepsilon_5>0$ accounts for the spatially uniform defocusing quintic nonlinearity.

Solitons solutions are looked for as $q=u(\eta)\exp(ib\xi)$, where $b$ is the propagation constant, and function $u$ represents the shape of the dissipative soliton pinned to the localized gain:

$$\frac{1}{2}\frac{d^2u}{d\eta^2}-bu+|u|^2 u+i\gamma u-i\Gamma\exp\left(-\frac{\eta^2}{w_\Gamma^2}\right)|u|^2 u-\varepsilon_5|u|^4 u=0\,. \tag{2}$$

As follows from here, stationary modes satisfy the condition of the balance between the gain and loss: $\Gamma\int_{-\infty}^{+\infty}\exp\left(-\eta^2/w_\Gamma^2\right)|u(\eta)|^4 d\eta=\gamma\int_{-\infty}^{+\infty}|u(\eta)|^2 d\eta\equiv\gamma U$, where $U$ is the total power.

First, we present analytical results based on the perturbation theory. To this end, assuming that width $w_\Gamma$ of the HS is much smaller than the size of the soliton pinned to it, we approximate the gain profile by $\tilde{\Gamma}\delta(\eta)$, with $\tilde{\Gamma}\equiv\sqrt{\pi}w_\Gamma\Gamma$ (so as to keep the same integral gain), and treat $\Gamma$ and $\gamma$ as small parameters. In the zero-order approximation, we take the exact soliton solution to Eq. (2) with $\gamma=\Gamma=0$ [29]:

$$u^2(\eta)=\frac{4\sqrt{b_{\rm co}}\,b}{\sqrt{b_{\rm co}-b}\cosh\left(2\sqrt{2b}\eta\right)+\sqrt{b_{\rm co}}}\,, \tag{3}$$

where the propagation constant and peak power are limited by the cutoff values, $0<b<b_{\rm co}\equiv 3/(16\varepsilon_5)$, $u^2(\eta=0)<P_{\rm co}\equiv u^2(\eta=0,b=b_{\rm co})=4b_{\rm co}$, and the total power is

$$U = \sqrt{8b_{\rm co}}\ln\left(\frac{\sqrt{b_{\rm co}}+\sqrt{b}}{\sqrt{b_{\rm co}-b}}\right). \qquad (4)$$

Numerical results presented below suggest that stable solitons correspond to $b_{\rm co}-b \ll b_{\rm co}$. In this case, the substitution of Eq. (3) into the power-balance condition, $\tilde{\Gamma}u^4(\eta=0)=\gamma U$, yields, in the first approximation of the perturbation theory,

$$U = 9\sqrt{\pi}\Gamma w_\Gamma / \left(16\gamma\varepsilon_5^2\right). \qquad (5)$$

This result demonstrates the necessity of the defocusing quintic term, as the total power diverges at $\varepsilon_5 \to 0$. Further, the width of broad solitons with the nearly constant peak power is $W \equiv U/P_{\rm co} \approx \left(3\sqrt{\pi}/4\right)\Gamma w_\Gamma/\left(\gamma\varepsilon_5\right)$.

Without assuming $b_{\rm co}-b \ll b_{\rm co}$, the analysis of the balance condition shows that the solutions exist for the gain strength exceeding the threshold value:

$$\Gamma \geq \Gamma_{\rm thr} = C\gamma\varepsilon_5^{3/2}/w_\Gamma, \qquad (6)$$

with numerical constant $C \approx 4.383$. At the threshold point, the (upper) soliton branch (5) is connected to a lower one, which can be found explicitly at $b \ll b_{\rm co}$, i.e., $w_\Gamma\Gamma \gg 5\varepsilon_5^{3/2}\gamma$:

$$U \approx \left(\frac{16\gamma}{\sqrt{\pi}w_\Gamma\Gamma}\right)^{1/3},\ b \approx \left(\frac{\gamma}{\sqrt{2\pi}w_\Gamma\Gamma}\right)^{2/3}. \qquad (7)$$

The lower branch is obviously unstable, being a separatrix between the stable zero solution and the upper branch. Note that $\varepsilon_5$ does not appear in solution (7), hence it remains valid, as the unstable one, in the absence of the quintic nonlinearity.

*Numerical results*. Figures 1 and 2 demonstrate that numerical findings confirm predictions of the above analysis. Soliton solutions exist if the gain exceeds the threshold value, as predicted by Eq. (6). Two branches of the soliton solutions originate from the threshold point, the upper one being almost exactly predicted by Eq. (5), while the lower branch is close to asymptotic approximation (7). In fact, a numerical solution of the perturbative balance equation yields curves which are virtually identical to those displayed in Figs. 1 and 2 (unlike the dotted lines that demonstrate some discrepancy), but such perturbative results cannot be presented in an explicit analytical form, on the contrary to Eqs. (5) and (7).

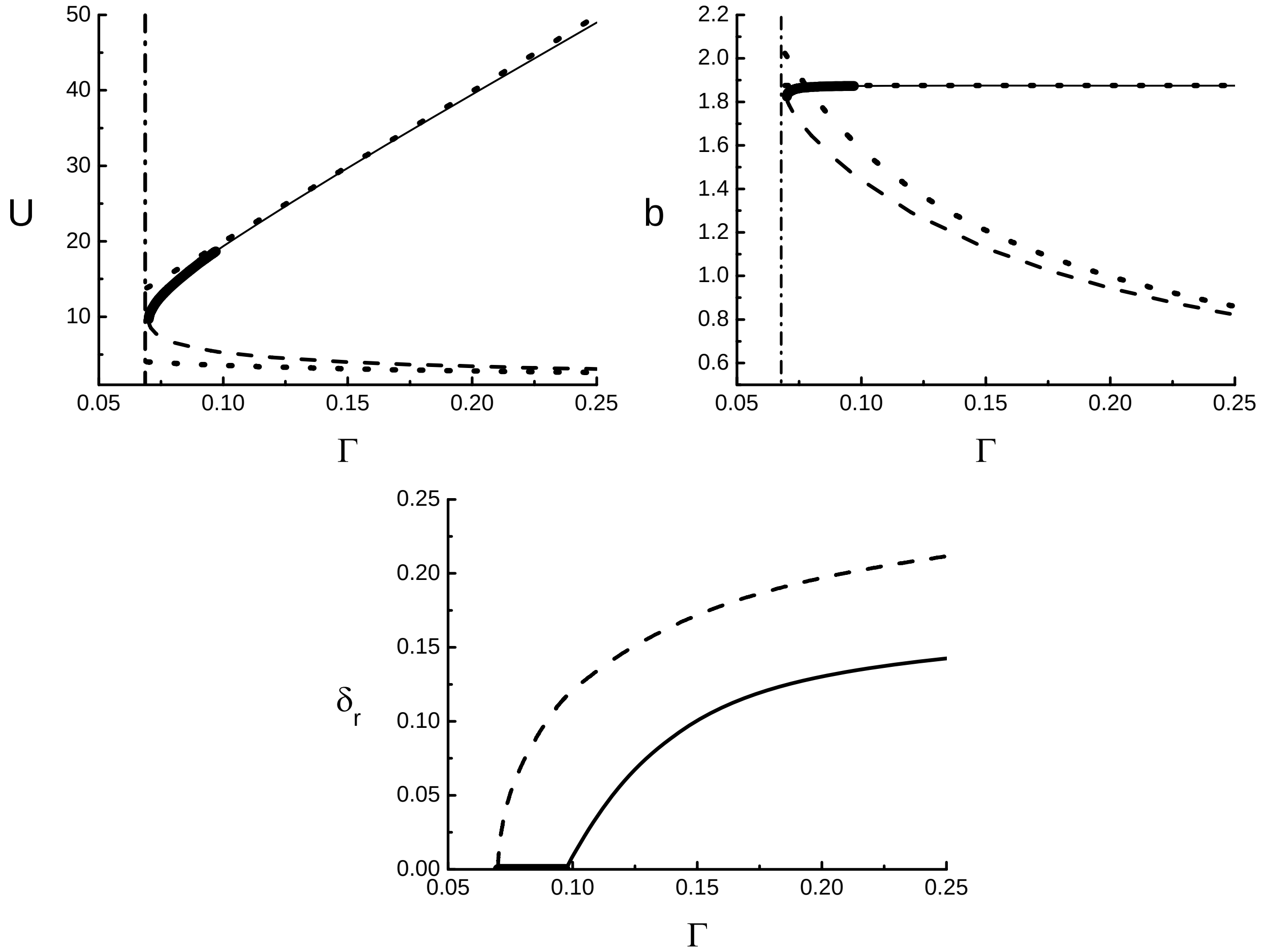

Fig. 1. The soliton's total power $U$ (top left), propagation constant $b$ (top right), and instability growth rates (bottom) versus gain coefficient $\Gamma$ at $\varepsilon_5 = 0.1$, $\gamma = 0.05$, $w_\Gamma = 0.1$. Here and in Fig. 2 below, the dashed line corresponds to the completely unstable lower branch, the bold segment represents the stable part of the upper branch, and its unstable continuation is depicted by the thin solid line. Analytical predictions (5) and (7) are shown by dotted curves (for the upper branch in the top right panel, the approximation amounts to $b = b_{co}$), and vertical dashed-dotted lines marks the threshold point as predicted by Eq. (6).

While the solitons belonging to the lower branch are completely unstable (as said above), suffering a rapid decay in the course of the propagation, a part of the upper branch is stable, as verified by the linear-stability analysis and direct simulations alike. The fully unstable lower branch is characterized by a real instability growth rate, while the destabilization of the higher branch at the border between the bold and thin curves in Figs. 1 and 2 is accounted for by a pair of complex eigenvalues (the Hopf bifurcation).

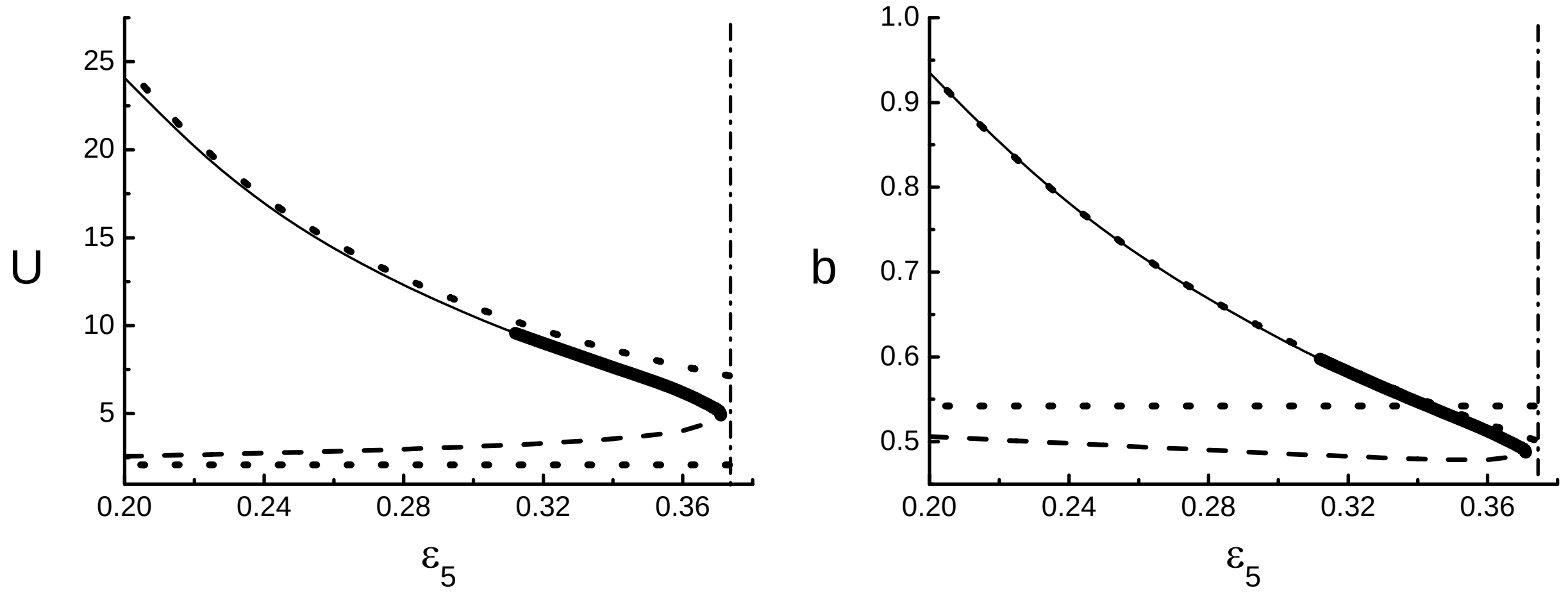

Fig. 2. The total power and propagation constant of the soliton versus the quintic coefficient, $\varepsilon_5$, at $\Gamma = 0.5$, $\gamma = 0.05$, $w_\Gamma = 0.1$. The horizontal dotted lines represent asymptotic values (7) which do not depend on $\varepsilon_5$.

Accordingly, in direct simulations the unstable soliton belonging to the upper branch is transformed into a stable localized breather (Fig. 3(a)). At still larger values of $\Gamma$, the breathers too become unstable, decaying in the course of the propagation. It is worthy to stress that, somewhat counter-intuitively, the increase of the cubic gain ($\Gamma$) leads to the decay of the breathers, rather than their blowup. This may be explained by the fact that the onset of the instability causes the expansion of the wave field into the lossy medium, where it suffers the attenuation, see Fig. 3(b).

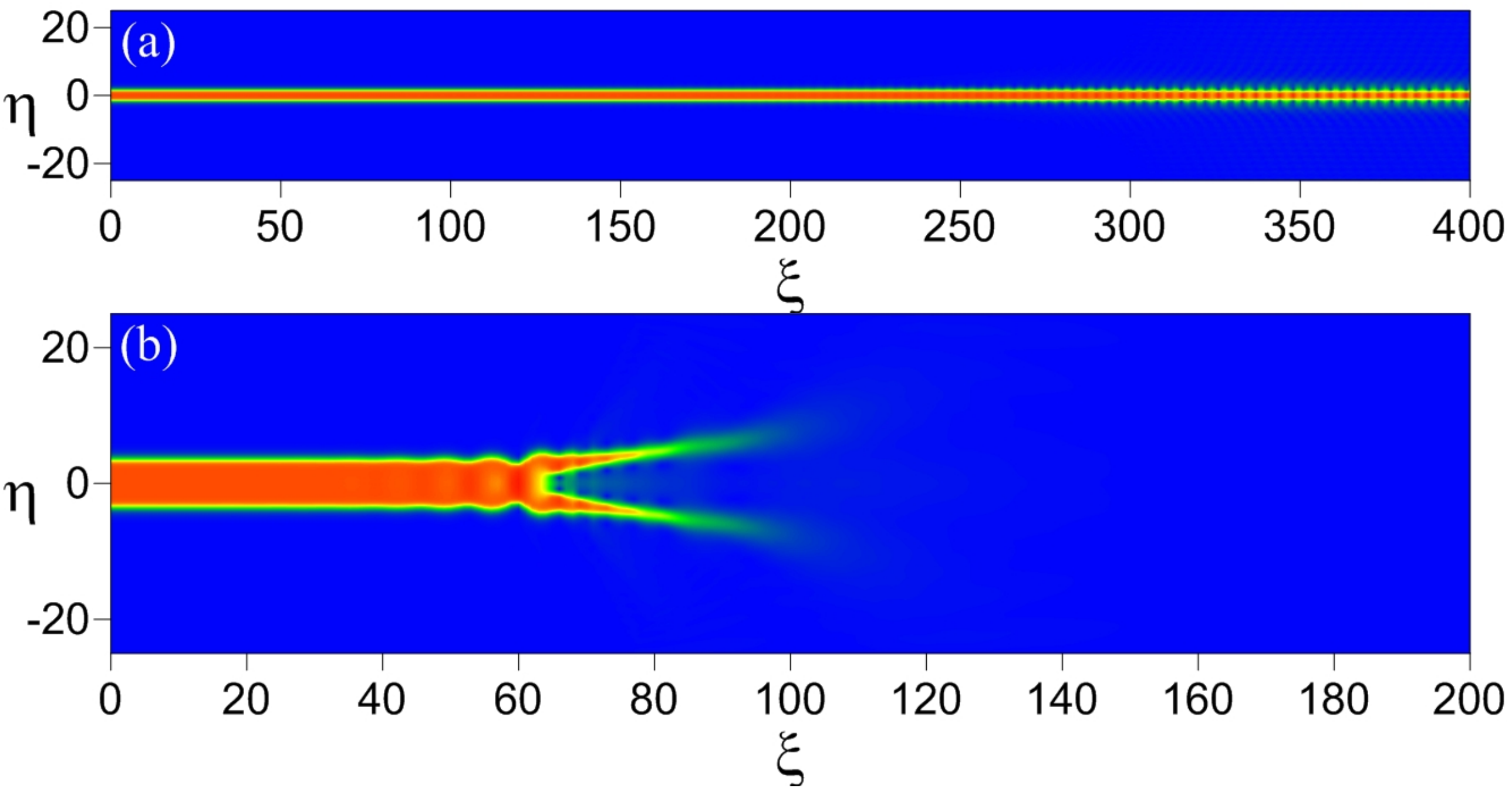

Fig. 3 (Color online) (a) The evolution of a weakly unstable soliton, belonging to the upper branch, into a stable breather, at $\varepsilon_5 = 0.1$, $\gamma = 0.05$, $w_\Gamma = 0.1$, and $\Gamma = 0.105$. (b) The decay of the localized mode at a larger value of the cubic gain, $\Gamma = 0.25$.

If the gain is fixed, while the quintic coefficient, $\varepsilon_5$, is varied, Fig. 2 demonstrates that the solitons exist below a maximum value of $\varepsilon_5$, which is also accurately predicted by Eq. (6): $\varepsilon_5 \leq \left[\Gamma w_\Gamma / \left(C\gamma\right)\right]^{2/3}$. In agreement with Eq. (5), $U$ is a monotonously decreasing function of $\varepsilon_5$ on the upper branch, and the respective dependence $b(\varepsilon_5)$ is close to $b_{co}(\varepsilon_5)$ (recall that $b \approx b_{co}$ along the upper branch).

Generic examples of numerically found soliton profiles are displayed in Fig. 4. The solitons belonging to the upper branch broaden with the increase of $\Gamma$, as predicted by the above analysis.

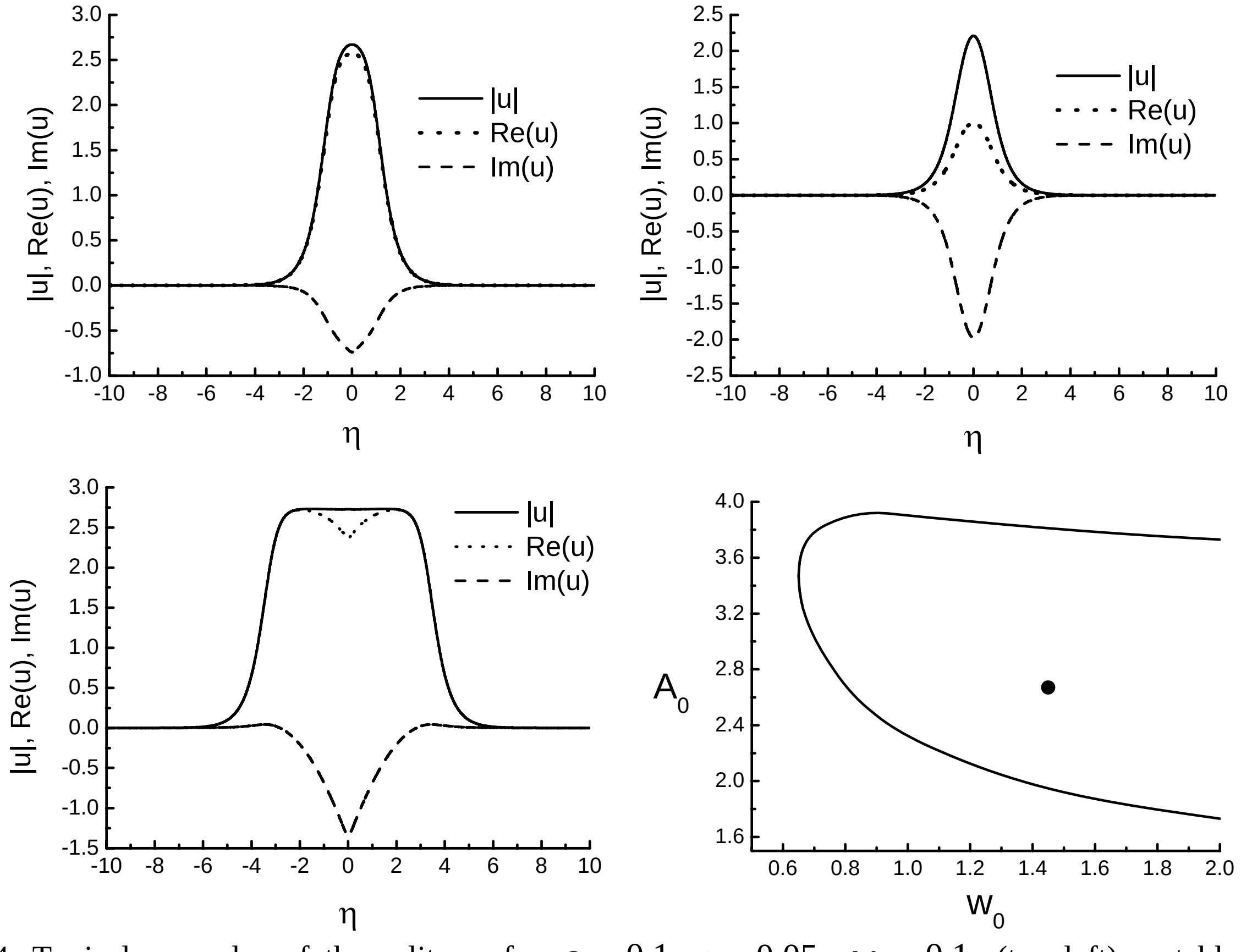

Fig. 4. Typical examples of the solitons, for $\varepsilon_5 = 0.1$, $\gamma = 0.05$, $w_\Gamma = 0.1$: (top left) a stable soliton belonging to the upper branch, at $\Gamma = 0.08$; (bottom left) a destabilized soliton from the same branch, at $\Gamma = 0.25$; (top right) an unstable soliton which belongs to the lower branch, at $\Gamma = 0.08$. The bottom right panel displays the attraction basin of the stable soliton excited by the Gaussian input (see text), the bold point showing parameters of the soliton fitted to the Gaussian shape.

Due to the bistability, i.e., competition with the stable zero solution, simulations demonstrate that the input in the form of the Gaussian, $u(\eta) = A_0 \exp\left(-\eta^2 / W_0^2\right)$, evolves into a stable soliton within a certain attraction domain in the plane of $\left(W_0, A_0\right)$, see an example in Fig. 4 (in models with the localized linear gain, any input evolves into the soliton [17-25]). Below the bottom border of the attraction domain, the input rapidly decays, while far above the top border it blows up (the latter outcome would be staved off by the quintic loss, i.e., saturation of the gain).

The results obtained for the solitons are summarized in the form of stability domains in the $\left(\Gamma, \varepsilon_5\right)$ plane, plotted in Fig. 5. Solitons do not exist above the top stability border, and they are unstable below the bottom border. With the increase of the loss parameter, $\gamma$, the stability domains shift to smaller values of $\varepsilon_5$. On the other hand, they shift to larger values of $\varepsilon_5$ and become narrower with the increase of width $w_\Gamma$ of the gain profile. These trends are explained by the fact that the integral gain, which is proportional to $\Gamma w_\Gamma$, must compensate the total loss rate, which is proportional to $\gamma$ and the width of the soliton. In turn, the soliton broadens with the increase of the strength of the defocusing quintic nonlinearity, $\varepsilon_5$, as shown above. Accordingly, at equal values of the loss and integral gain, the values of $\varepsilon_5$ for stable solitons are nearly equal too, see the right panel in Fig. 5. The fact that the stability regions are narrow in terms of $\Gamma$ is not a problem for the experimental realization, as the gain may be easily varied by adjusting the pump intensity.

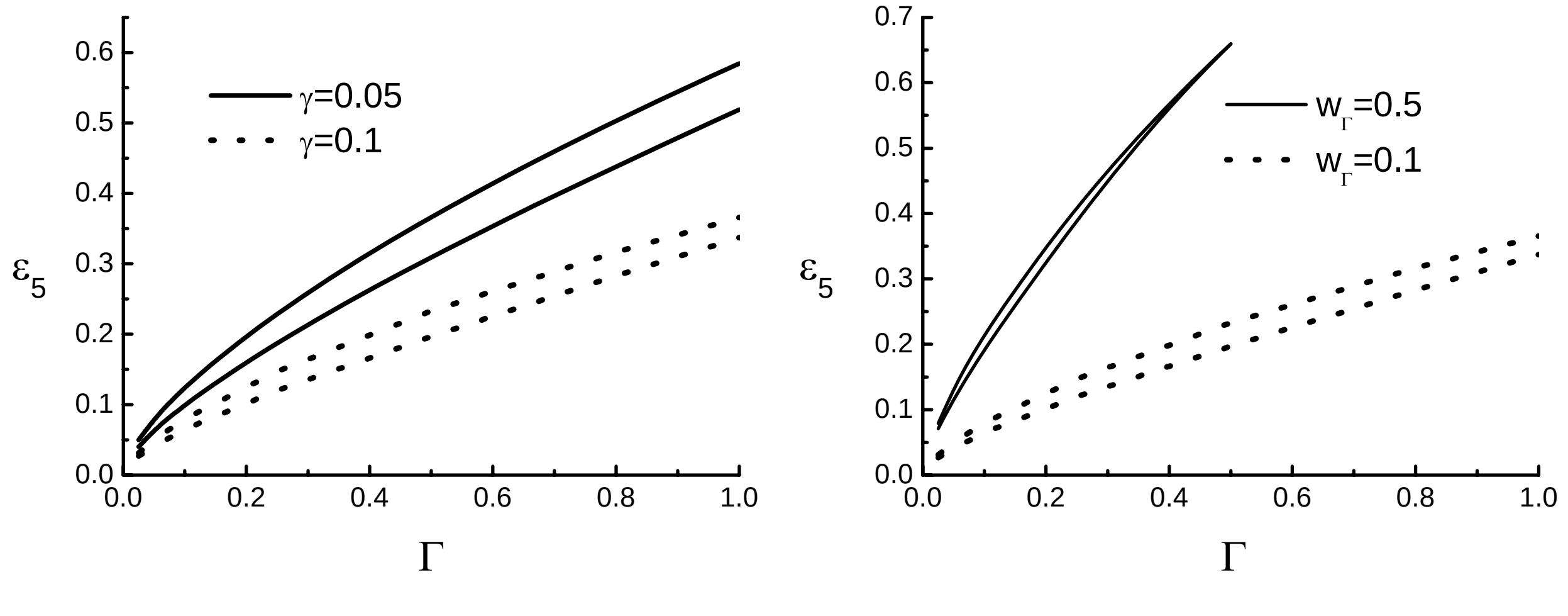

Fig. 5. Stability domains (between the borders) for the solitons at different values of the loss coefficient, $\gamma$, and fixed width of the gain distribution, $w_{\Gamma} = 0.1$ (the left panel), and at different values of $w_{\Gamma}$ and fixed $\gamma = 0.1$ (the right panel).

*Conclusion*. We have introduced the model featuring the localized cubic gain (alias HS, "hot spot") embedded into the medium with the cubic-quintic nonlinearity and linear loss. Using the perturbation theory and systematic numerical analysis, we have found a family of fundamental spatial solitons pinned to the HS. On the contrary to the CGL equation modeling uniform media, a part of the family is stable without the overall-stabilizing quintic dissipation, a necessary ingredient being the defocusing quintic nonlinearity. The family of the stable solitons is bounded by the Hopf bifurcation, which replaces them by breathers. Eventually, the breathers lose their stability too and decay.

This setting may be used as a nonlinear amplifier for optical and plasmonics solitons which does not require the gain saturation (accounted for by the quintic losses). It may be interesting to extend the analysis for higher-order solitons, in the case of a sufficiently broad HS. A challenging issue is to analyze a two-dimensional version of the model.

We appreciate valuable discussions with Y.V. Kartashov and L. Torner. The work of O.V. Borovkova was supported by the Ministry of Science and Innovation, Government of Spain, grant FIS2009-09928.

## REFERENCES

[1] Lederer F., Stegeman G. I., Christodoulides D. N., Assanto G., Segev M., and Silberberg Y., Phys. Rep. **436** (2008) 1.

[2] Kartashov, Y. V., Vysloukh V. A., and Torner L., in *Progress in Optics* (Elsevier B. V., Amsterdam, 2009), Vol. 52, p. 63.

[3] Kartashov Y. V., Malomed B. A., and Torner L., Rev. Mod. Phys. **83** (2011) 247.

[4] Rosanov, N. N., *Spatial Hysteresis and Optical Patterns* (Springer: Berlin, 2002).

[5] Malomed B. A. and Winful H. G., Phys. Rev. E **53** (1996) 5365.

[6] Atai J. and Malomed B. A., Phys. Rev. E **54** (1996) 4371.

[7] Paulau P. V., Gomila D., Colet P., Loiko N. A., Rosanov N. N., Ackemann T. and Firth W. J., Opt. Exp. **18** (2010) 8859.

[8] Malomed B. A., Physica D **29** (1987) 155.

[9] van Saarloos W. and Hohenberg P. C., Phys. Rev. Lett. **64** (1990) 749.

[10] Hakim V., Jakobsen P., and Pomeau Y., Europhys. Lett. **11** (1990) 19.

[11] Malomed B. A. and Nepomnyashchy A. A., Phys. Rev. A **42** (1990) 6009.

[12] Marcq P., Chaté H., and Conte R., Physica D **73** (1994) 305.

[13] Matsumoto M., Hasegawa A., and Kodama Y., Opt. Lett. **19** (1994) 1019.

[14] Gabitov I., Holm D., Luce B., and Mattheus A., Opt. Lett. **20** (1995) 2490.

[15] Burtsev S., Kaup D. J., and Malomed B. A., J. Opt. Soc. Am. B **13** (1996) 888.

[16] Chu P. L., Malomed B. A., and Peng G. D., Opt. Commun. **128** (1996) 76.

[17] Lam C.-K., Malomed B. A., Chow K. W., and Wai P. K. A., Eur. Phys. J. Special Topics **173** (2009) 233.

[18] Kartashov Y. V., Konotop V. V., and Vysloukh V. A., EPL **91** (2010) 340003.

[19] Tsang C. H., Malomed B. A., Lam C.-K., and K. W. Chow, Eur. Phys. J. D **59** (2010) 81.

[20] Zezyulin D. A., Kartashov Y. V., and Konotop V. V., Opt. Lett. **36** (2011) 1200.

[21] Kartashov Y. V., Konotop V. V., and Vysloukh V. A., Phys. Rev. A **83** (2011) 041806(R).

[22] Tsang C. H., Malomed B. A., and Chow K. W., Phys. Rev. E **84** (2011) 066609.

[23] Lobanov V. E., Kartashov Y. V., Vysloukh V. A., and Torner L., Opt. Lett. **36** (2011) 85.

[24] Borovkova O. V., Lobanov V. E., Kartashov Y. V., and Torner L., Opt. Lett. **36** (2011) 1936.

[25] Mak W. C. K., Malomed B. A., and Chu P. L., Phys. Rev. E **67** (2003) 026608.

[26] Caboche E., Pedaci F., Genevet P., Barland S., Giudici M., Tredicce J., Tissoni G., Lugiato L.A., Phys. Rev. Lett. **102** (2009) 163901.

[27] Schöpf W. and Kramer L., Phys. Rev. Lett. **66** (1991) 2316.

[28] Kim S., Jin J. H., Kim Y. J., Park I. Y., Kim Y., and Kim S. W., Nature **453** (2008) 757.

[29] Pushkarov Kh. I., Pushkarov D. I., and Tomov I. V., Opt. Quant. Electr. **11** (1979) 471.